\documentclass[letterpaper]{article} 
\usepackage{aaai25}  
\usepackage{times}  
\usepackage{helvet}  
\usepackage{courier}  
\usepackage[hyphens]{url}  
\usepackage{graphicx} 
\usepackage{natbib}  
\usepackage{caption} 
\usepackage{algorithm}
\usepackage{algorithmic}
\usepackage{multicol}
\usepackage{multirow}
\usepackage{amsmath}
\usepackage{amssymb}
\usepackage{bm}
\usepackage{tikz}
\usepackage{subcaption}

\usepackage{newfloat}
\usepackage{listings}
\DeclareCaptionStyle{ruled}{labelfont=normalfont,labelsep=colon,strut=off} 
\floatstyle{ruled}
\newfloat{listing}{tb}{lst}{}
\floatname{listing}{Listing}
\title{Auto-Regressive Diffusion for Generating 3D Human-Object Interactions}
\author{
    Zichen Geng\textsuperscript{\rm 1},
    Zeeshan Hayder\textsuperscript{\rm 2},
    Wei Liu\textsuperscript{\rm 1},
    Ajmal Saeed Mian  \textsuperscript{\rm 1 \thanks{Corresponding author}} 
}
\affiliations{

    \textsuperscript{\rm 1}The University of Western Australia\\
    35 Stirling Highway, \\
    Perth, WA 6009 Australia\\

    \textsuperscript{\rm 2}Commonwealth Scientific and Industrial Research Organization\\
    Synergy building Black Mountain \\
    Canberra, ACT 2601 Australia\\
    zen.geng@research.uwa.edu.au,
    zeeshan.hayder@data61.csiro.au
    wei.liu@uwa.edu.au,
    ajamal.mian@uwa.edu.au

}

\usepackage{bibentry}

\begin{document}

\maketitle

\begin{abstract}
Text-driven Human-Object Interaction (Text-to-HOI) generation is an emerging field with applications in animation, video games, virtual reality, and robotics. 
A key challenge in HOI generation is maintaining interaction consistency in long sequences. 
Existing Text-to-Motion-based approaches, such as discrete motion tokenization, cannot be directly applied to HOI generation due to limited data in this domain and the complexity of the modality. 
To address the problem of interaction consistency in long sequences, we propose an autoregressive diffusion model (ARDHOI) that predicts the next continuous token.
Specifically, we introduce a Contrastive Variational Autoencoder (cVAE) to learn a physically plausible space of continuous HOI tokens, thereby ensuring that generated human-object motions are realistic and natural. 
For generating sequences autoregressively, we develop a Mamba-based context encoder to capture and maintain consistent sequential actions. Additionally, we implement an MLP-based denoiser to generate the subsequent token conditioned on the encoded context. Our model has been evaluated on the OMOMO and BEHAVE datasets, where it outperforms existing state-of-the-art methods in terms of both performance and inference speed. This makes ARDHOI a robust and efficient solution for text-driven HOI tasks.

\end{abstract}


\begin{links}
     \link{Code}{https://github.com/gengzichen/ARDHOI}
\end{links}

\section{Introduction}

Text-driven Human-Object Interaction (Text-to-HOI) is a derived task of Text-to-Motion. Text-to-Motion aims to synthesize realistic human motion sequences with text prompts, while Text-to-HOI generates both human motions and object motions. Generating realistic interactions has extensive applications in animation, video games, film-making, virtual reality, embodied intelligence \cite{physhoi}, and robotics \cite{embodied1}.

\begin{figure}[htp]
    \centering
    \includegraphics[width=\linewidth]{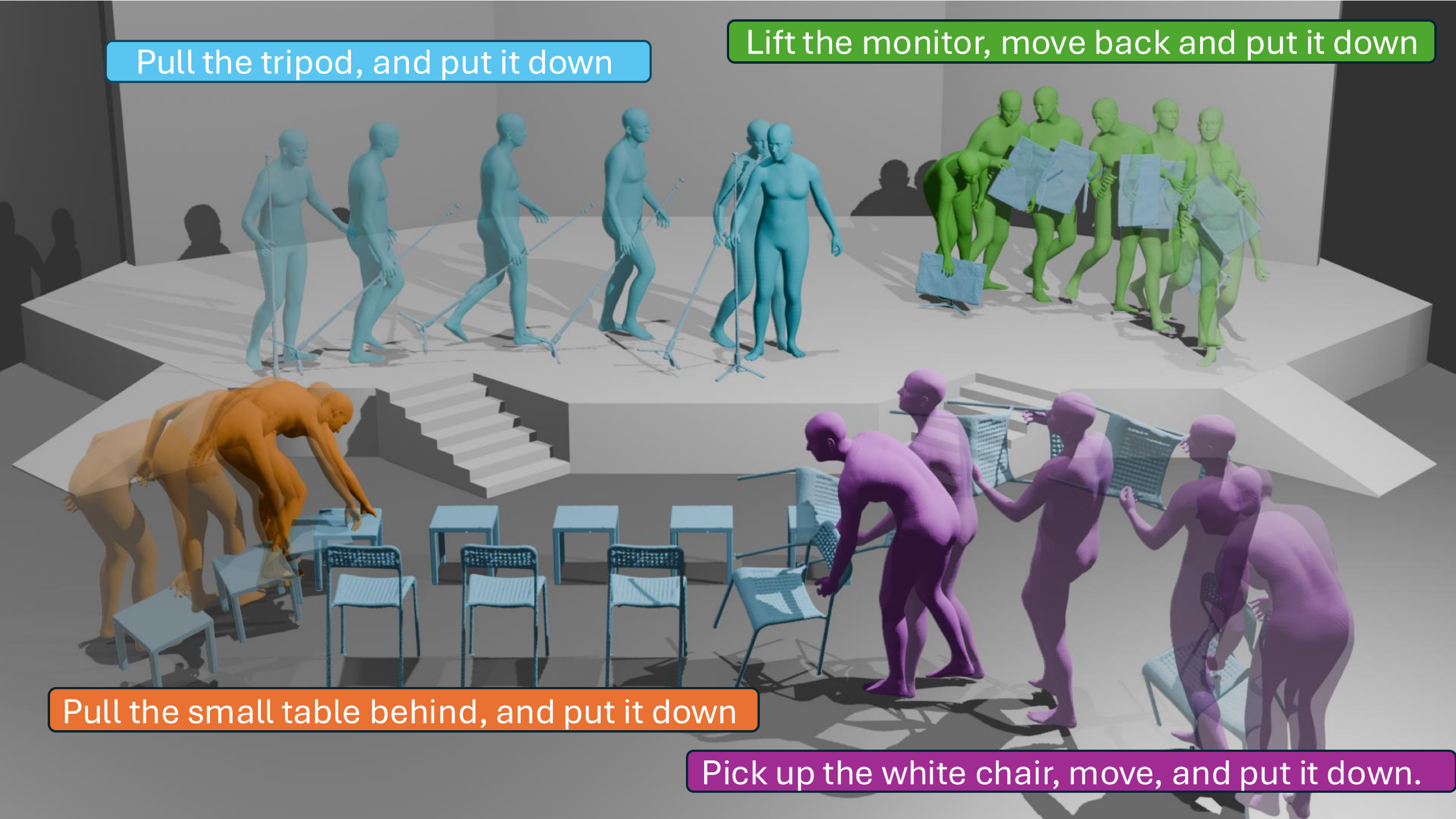}
    \caption{Text guided Human-Object interaction motion sequences generated by our ARDHOI.} 
    \label{fig:main}
\end{figure}

Text-to-Motion has been gaining significant traction in the research community lately, thanks to the development of universal network architectures like Transformer \cite{vaswani2017attention} and generative algorithms like the diffusion model \cite{ddpmo, ddpm}. 

The success of generative paradigms in Text-to-Motion tasks has also driven significant advancements in related tasks, such as multi-person motion generation \cite{intergen} and motion prediction \cite{interdiff, multipersonprediction}. These paradigms primarily include diffusion-based and autoregressive models. Among them, autoregressive models, such as T2M-GPT \cite{t2mgpt}, AttT2M \cite{attt2m}, MotionGPT \cite{motiongpt}, MoMask \cite{guo2024momask} have achieved state-of-the-art performance in Text-to-Motion. 

The success of autoregressive models 
is driven by the consistent nature of the teacher-forcing mechanism, which explicitly supervises the prediction of subsequent positions in a sequence. In contrast, diffusion-based models have no such contextual awareness. Additionally, diffusion-based models typically require the sequence length (either learned or provided as ground truth) to truncate the sequence to a desired duration, while autoregressive models can intrinsically determine when the sequence should end. This feature streamlines the generation process and improves efficiency. 

We believe that such autoregressive mechanisms can also benefit Text-to-HOI. However, directly migrating previous paradigms and architectures to Text-to-HOI is challenging.  Firstly, conventional autoregressive motion generative models such as T2M-GPT \cite{t2mgpt} and Motion-GPT \cite{motiongpt} are highly dependent on discrete tokenization, where each next token is predicted by optimizing the cross entropy. Such tokenization is typically facilitated through a Vector Quantized Variational Autoencoder (VQ-VAE) which encodes small motion intervals into a quantized codebook, compressing continuous motion into a discrete space. While this pre-process approach maintains the coherence and realism of generated motions over extended periods, it falls short when applied to smaller datasets, as highlighted by \citet{ardiff}. The quantization process inherent in VQ-VAE limits generalization, particularly when the HOI training data is small \cite{omomo, behave}, where contact patterns cannot be fully learned from limited samples. 

Another challenge arises from the interplay between human motion and object motion. Even when physical constraints like contact and collision are used to model interactions between humans and objects, generating a physically plausible sequence using the forward method (without post-processing optimization or guidance) remains challenging. To address the inconsistency issues in human-object interactions and to fully leverage the benefits of autoregressive architecture without quantizing HOI tokens, we propose an autoregressive model that performs diffusion in the continuous token space. Additionally, to enhance inference speed, we utilize the Mamba state space model \cite{mamba} to capture the contextual information of previous tokens.


Given the limitations of discrete tokens in representing HOI, continuous tokens offer a more suitable alternative. Variational Autoencoders (VAEs) are commonly used in conventional continuous token space learning due to their adaptability \cite{rombach2021highresolution}. However, na\"ive VAE training can lead to overfitting on small datasets, resulting in rigid motion patterns that do not generalize. This highlights the need for learning a more comprehensive latent space for robust HOI representation. Furthermore, the dependency of human and object motions complicates the explicit modeling of the physical plausibility of contact and motion in HOI data within the original 3D space.
To such an extent, a na\"ive VAE-based tokenizer, without proper constraints, may yield physically implausible HOI representations. 
For example, in 3D space, a small perturbation in object translation may seem insignificant but can appear highly implausible to human observers. Hence, when the two HOI samples are mapped into a latent space of a na\"ive VAE, they tend to be very close to each other, as shown in Fig.~\ref{fig:PCA_Comparison}.


To tackle these issues and develop an effective contact-aware latent space, we introduce a Contrastive Variational Autoencoder (cVAE). This model learns a continuous token space by enhancing plausible samples (based on certain criteria) and widening the gap between plausible and implausible HOI tokens. This approach ensures that implausible HOI tokens that are close to plausible ones in 3D space are not mapped to nearby positions in the latent token space. Furthermore, the clear distinction between positive and negative token spaces helps minimize error accumulation during autoregressive processes.


Our approach differs from previous HOI models \cite{interdiff, hoidiff, thor} in several ways. \citet{interdiff, hoidiff} use pre-trained modules for classifier guidance in the inverse diffusion process to ensure interaction awareness. However, \citet{thor} employs predictors for post-processing contact optimization of denoised human motion. In contrast, our method focuses on the prior distribution of HOI tokens. Instead of constraining interactions across the entire sequence, we constrain plausible HOI clips within the latent space. This approach reduces computational requirements during the reverse diffusion process.
Our contributions and findings are summarized as follows:

\begin{itemize}
    \item We propose the Autoregressive Diffusion for Human-Object Interaction (ARDHOI), a versatile paradigm for Text-to-HOI generation that utilizes only Mamba and MLP, operating in a forward prior manner without the need for post-optimization.

    \item We propose a contrastive-learning-based VAE (cVAE) that tokenizes the HOI sequence while incorporating kinematical constraints and object interaction awareness. This approach addresses the issues of insufficient HOI prior knowledge and reduces error accumulation during autoregression.

    \item We propose an autoregressive diffusion method in the continuous token space to produce contextually consistent and high-fidelity HOI sequences. Our approach effectively leverages comprehensive context information. Experiments indicate that our Mamba-based context encoder captures contextual details more effectively than Transformer-based architectures.

    \item We showcase the effectiveness of the MLP denoiser in the reverse diffusion process, highlighting its ability to deliver both high inference speed and superior accuracy compared to advanced cross-attention methods.


    
    
\end{itemize}

Experiments on the OMOMO \cite{omomo} and BEHAVE \cite{behave} datasets demonstrate that our method outperforms current SOTA techniques in both accuracy and inference speed. For instance, on the OMOMO dataset, we achieve a 23\% reduction in FID error compared to the closest competitor, HOI-Diff \cite{hoidiff}.


\begin{figure*}
    \centering
    \includegraphics[width=0.95\linewidth]{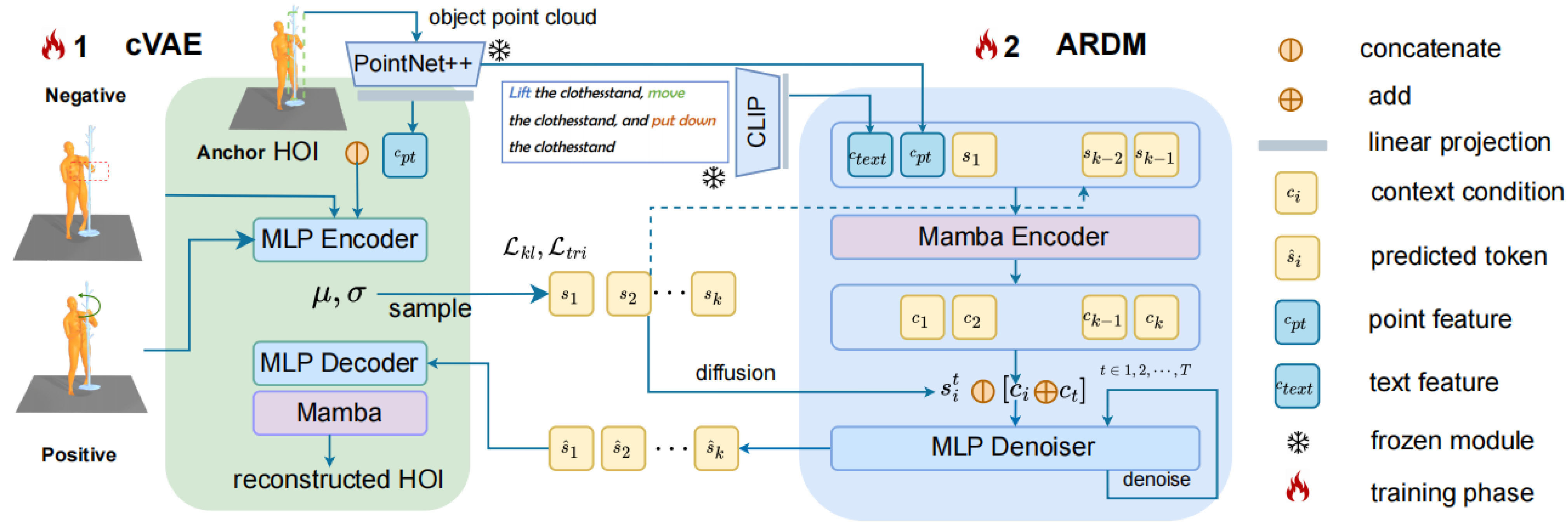} 
    
    \caption{ARDHOI. The left model is our Contrastive VAE (cVAE) which learns continuous HOI tokens in a contrastive manner. It is trained in phase 1 and frozen in phase 2.  The right part is the ARDM, which generates HOI tokens in an autoregressive style using diffusion. ARDM is trained in phase 2.}
    \label{fig:pipeline}
\end{figure*}

\section{Related Work}

\noindent \textbf{Text-to-Motion} has garnered significant attention with the rise of diffusion models \cite{ddpmo} and the establishment of foundational baselines in the field \cite{a2m}. Early methods primarily utilized VAE-based approaches \cite{a2m, temos} and GANs \cite{xu2023actformer}, but recent advancements have led to the development of two main categories of motion generative models: diffusion-based models and autoregressive models.
Diffusion-based models were first applied to the Text-to-Motion task by \citet{mdm, mofusion, motiondiffuse, kim2022flame}, who leveraged Transformers and diffusion models to operate directly on raw motion sequences. \citet{mld} further advanced this approach by introducing the first method to perform diffusion in latent space, utilizing a Transformer-based VAE. In the realm of autoregressive approaches, \citet{t2mgpt} pioneered the use of a CNN-based VQ-VAE in combination with GPT to predict discrete motion tokens. Subsequent works \cite{motiongpt, guo2024momask, attt2m} built upon this vector quantization approach, enhancing the autoregressive method. Some research has explored hybrid architectures that combine autoregression and diffusion algorithms. For instance, \citet{armd} introduced a cross-attention-based denoiser to predict the next frame based on the previous frame and text conditions, while \citet{cmdm} proposed a pure MLP-based autoregressive denoiser for synthesizing long motion sequences. However, these autoregressive models typically condition predictions only on previous frames and initial conditions, limiting their ability to fully capture contextual information. Moreover, they do not adequately address the problem of error accumulation during autoregressive training. To overcome these limitations, \citet{ardiff} recently proposed a new paradigm for image generation using continuous token space, bypassing the need for vector quantization. Their approach predicts the token distribution in continuous space through autoregressive diffusion algorithms, which better leverages contextual information throughout the sequence.

\noindent \textbf{Text-Driven Human-Object Interaction Generation}. Compared to the Text-to-Motion task, motion generation for human-object interaction (HOI) remains underexplored. Earlier works \cite{humanise, GTAIM, playforbenchmark, PROX} primarily focus on HOI in static scenes or static HOI actions \cite{posa, MIME}. Additionally, some research has utilized reinforcement learning to generate human motion in HOI tasks \cite{physhoi}. \citet{interdiff} introduced the first autoregressive diffusion-based approach for the HOI task, but their method conditions on historical HOI sequences rather than text. \citet{cghoi} and \citet{hoidiff} are among the earliest to propose diffusion models specifically for interactions between humans and dynamic objects. \citet{cghoi} guides the diffusion process by predicting contact data at each time step, while \citet{hoidiff} emphasizes the significance of affordance information in coarse-grained contact, introducing a module to generate affordances and performing optimization after the inverse diffusion process. Building on these foundations, \citet{thor} proposed a relation intervention-based diffusion model that refines object rotation during each inverse diffusion step. In a different vein, \citet{chois} presented an autoregressive approach to predict the next frame based on text, mesh, and an initial state. Recently, \citet{interdreamer} introduced InterDreamer, a World Model 
that leverages detailed textual planning generated by a Large Language Model and pre-trained Text-to-Motion models to achieve zero-shot HOI generation.

\noindent \textbf{Mamba} \cite{mamba, mamba2} is the latest state space model (SSM) and has demonstrated success across various domains, including natural language processing, computer vision \cite{mambasuite, visionmamba}, and motion generation. Mamba employs a convolutional neural network (CNN) architecture to enable low-memory-cost parallel training and rapid autoregressive inference. While most prior works rely on the Transformer architecture for its broad applicability in sequence tasks, \citet{motionmamba} introduced Motion Mamba, claiming it to be more effective for generating long sequences. However, Motion Mamba primarily functions as a denoiser, building on \cite{mld}'s Transformer VAE to map a 200-frame sequence into a $2\times256$ latent space. The model's role is limited to denoising the small global latent feature, without directly processing long sequences in a comprehensive manner.

\section{Text-driven HOI Generation}

Given a text description, and point cloud of object, 
our model aims to generate a human-object interaction sequence $\bm{x}$.
We present an autoregressive generative model in continuous latent space with two stages: autoregressive HOI generation in diffusion style and contrastive VAE for continuous HOI tokens.
Fig. ~\ref{fig:pipeline} illustrates our proposed approach.

\subsection{Contrastive VAE for HOI Token}

Previous approaches to Text-to-HOI incorporate additional information to guide the generative objectives or apply after-effect spatial constraints to refine the interaction \cite{hoidiff, cghoi}. Such post-correctness requires additional computations and fails to guarantee a meaningful motion, e.g. objects stuck to the limbs. In contrast, our approach focuses on learning an effective HOI latent space for short temporal intervals with contrastive learning.

VAEs can further refine the denoised sample generated by the diffusion model due to their denoising properties. \citet{mld} suggests that a Transformer-based VAE can learn a highly compressed (e.g. 1/100) latent representation of human-motion-only data. When combined with diffusion models, this latent space facilitates better reverse diffusion processing than raw motion data. Despite this, both the THOR model \cite{thor} and our experiments demonstrate that a Transformer-based VAE is not ideal for mapping entire HOI sequences into a latent space, even much less for generation purposes. The complexity of HOI data presents unique challenges: HOI datasets feature more intricate and fine-grained movement steps and contain fewer samples compared to human-only datasets like HumanML3D, which offer more sequences but fewer sub-movements. Additionally, traditional latent spaces are incapable of handling physically implausible contact scenarios.

To tackle the above issues, we propose a cVAE designed to learn contact-aware motion tokens through a contrastive learning approach. Positive and negative samples are constructed by adding small random offsets to object translations. An augmented sample is considered positive if its minimum and second minimum contact distances with the object increase by a certain threshold $\tau$ (for the same contact joints) after the offset. In this paper, we find that $\tau=3$cm is a reasonable value. Previous work \cite{coins} introduced a contrastive VAE in voxel space by adding simple Gaussian perturbations. In contrast, we found that for centrally symmetric objects, such as clothes stands, positive samples can be generated by applying random vertical rotations, which do not alter the contact situation. 
We utilize triplet loss \cite{triplet} to ensure that positive samples are mapped closer to the anchor samples in the latent space, while negative samples are kept distant.
\vspace{-2mm}

\begin{align}
& \mathcal{L}_{tri} = \max(||\bm{s}_i-\bm{s}_{i,p}||_2-||\bm{s}_i-\bm{s}_{i,n}||_2 + \alpha, 0),
\end{align}
\noindent where $\bm{s}_i$ is the $i$-th continuous token among the total $K$ tokens of the HOI sequence $\bm{x}$, $\bm{s}_{i,p}, \bm{s}_{i,n}$ are positive and negative tokens sampled from the latent space, and $\alpha > 0$ is the triplet margin. The margin $\alpha$ should be carefully set, neither too large nor too small. A small margin leads to high similarity of positive and negative samples while a high margin may result in an unbounded latent space. 
For motion representation, instead of the widely used HumanML3D \cite{tm2t} representation, we choose the SMPL representation \cite{SMPL}. The reason is that the HumanML3D representation is velocity-based, and global representation is obtained by integration from previous frames, which limits the global spatial representation at arbitrary frames. 

Our cVAE is lightweight, comprising only a few Multi-Layer Perceptrons (MLPs) and a single Mamba Layer. Motion sequences are first segmented into short frames (16 each) and flattened. Each flattened token is then concatenated 
with the point cloud feature of the object intended for interaction. Point cloud features are extracted using a pre-trained PointNet++ model \cite{pointnet++} trained for object classification. The MLP encoder maps these tokens to a Gaussian latent space, and the encoded motion tokens are subsequently fed into an MLP decoder. Since MLPs do not operate on the sequence dimension, we incorporate a single Mamba layer to enable token-wise smoothness and contextual reconstruction. In practice, we observed that Mamba intrinsically reduces jittering compared to Transformer-based models, likely due to its state space modeling. Apart from the primary reconstruction task of VAE, whose reconstruction loss is give by:
\vspace{-2mm}

\begin{align}
& \mathcal{L}_{rec} = \frac{1}{K}\sum_{i=1}^{K}||\bm{x}_i-\bm{\hat{x}}_{i}||_2,
\end{align}
\noindent where $K$ is the token number, and $\bm{x}_i$ is the $i$-th token. Besides, Kullback-Leibler Divergence is also added to the objective function to force the latent space to be compact.
\begin{align}
&  \mathcal{L}_{kl}=\sum_{i=1}^{K}\sum_{j=1}^{d_l} (\sigma_{i,j}^2 + \mu_{i,j}^2 -1 -\log{\sigma^2_{i,j}}),
\end{align}
\noindent where $\mu_{i,j}$, $\sigma_{i,j}$ are the mean and standard variance of the $i$-th token's $j$-th dimension and $d_l$ is the token number and dimension of the token space, respectively. Besides, physical constraints (see supplementaries. for details) are included to make the reconstructed result more physically plausible.
%
\begin{equation}
\mathcal{L}_{phy}= \lambda_{fk}\mathcal{L}_{fk} + \lambda_{vel}\mathcal{L}_{vel} +\lambda_{ovel}\mathcal{L}_{ovel} +\lambda_{con}\mathcal{L}_{con},
\end{equation}
\noindent where the $\mathcal{L}_{fk}$ is the L2 norm of forward kinematic joint position error, the $\mathcal{L}_{vel}$ is the L2 norm of joint velocity loss, $\mathcal{L}_{ovel}$ is the L2 norm of object motion velocity loss, and $\mathcal{L}_{con}$ is the contact loss, which is the mean of pair-wise difference between the ground truth contact distance and predicted contact distance. The overall objective function is:
%
%
\begin{align}
& \mathcal{L}_{cVAE} = \mathcal{L}_{rec} + \lambda_{tri}\mathcal{L}_{tri} + \lambda_{kl}\mathcal{L}_{kl} + \lambda_{phy}\mathcal{L}_{phy},
\end{align}
Our approach not only addresses the complexities inherent in HOI data but also enhances the generation of realistic and coherent human-object interactions, paving the way for more sophisticated applications in virtual environments.

\subsection{Autoregressive Diffusion Model}
Conventional diffusion-based models \cite{mdm, thor, cghoi, hoidiff} aim to predict the sample or noise on the whole sequence. However, such an approach overlooks the autoregressive nature of sequences, and increases the learning difficulty in long sequences, leading to inconsistency. To address this problem, we propose an Autoregressive Diffusion Model (ARDM) using continuous HOI tokens to generate long consistent sequences through teacher forcing. The
log-likelihood $\log p_{\theta}(s)$ of sequence $\bm{s}$ is factorized as:
%
\begin{equation}
    \log p_{\theta}(\bm{s}) = \sum_{i=1}^{K} \log p_{\theta}(\bm{s}_i | \bm{s}_{1:i-1}),
\end{equation}
where $\bm{s}_{1:i-1} := \{\bm{s}_1, \bm{s}_2, \cdots , \bm{s}_{i-1} \}$ is the previous HOI token preceding the $i$-th position.

\noindent \textbf{Mamba Context Encoder}. Our ARDM consists of two sub-networks: an autoregressive context encoder based on Mamba, and an MLP denoiser. The context encoder encodes the textual condition $c_{text}$ and object conditions $c_{pt}$, together with the previous motion tokens $c_{j}, j \in \{1, \cdots, i-1\}$ as conditions to predict the next token at position $i$. Since contact-aware features and physics constraints have been learned by the cVAE, no additional constraint is needed to add to the denoiser training. The encoded context condition is then added by projected diffusion time step condition $t$ independently on each token position. Notably, a major concern for autoregressive models is the accumulation of error. To avoid error accumulation, some models apply techniques such as masking and additional discriminators to predict the generation qualities of the next tokens. In Text-to-HOI, the accumulated error of human-object interaction gets amplified during the inference stage \cite{interdiff}. However, the proposed cVAE ensures that the negative samples are well separated from the positive ones with a contrastive margin, serving as an implicit discriminator. Therefore, the latent token space guarantees that the predicted token will not drift to an implausible token during the reverse diffusion.

\begin{figure}
    \centering
    \includegraphics[width=0.8\linewidth]{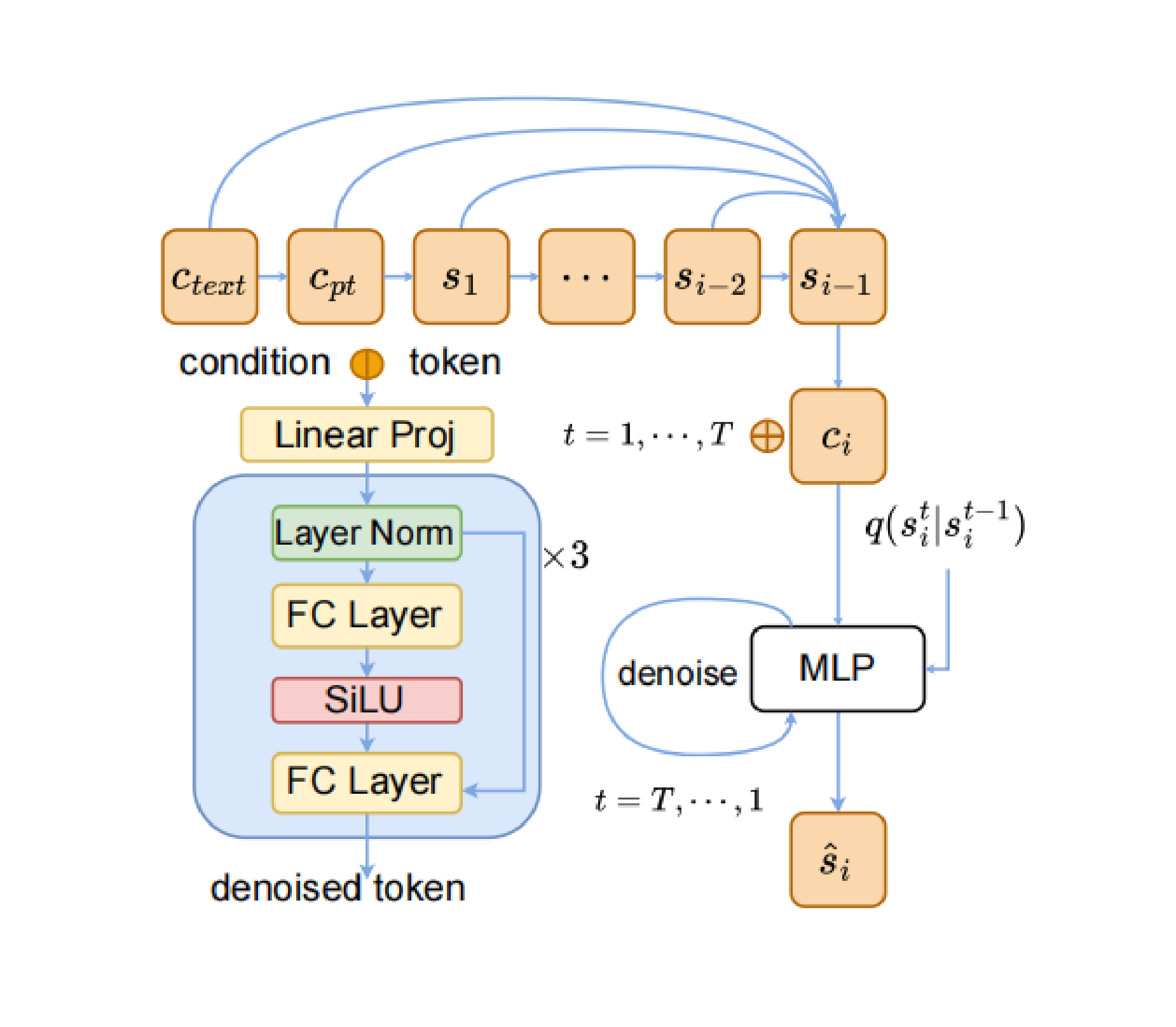}
    \vspace{-5mm}
    \caption{MLP Denoiser and autoregressive denoising}
    \vspace{-4mm}
    \label{fig:mlpdenoiser}
\end{figure}

\noindent \textbf{MLP Denoiser}. To implement the denoiser, instead of a heavy architecture such as a Transformer, we only employ a lightweight MLP for efficiency as shown in Table ~\ref{fig:mlpdenoiser}. The tokens with scheduled noise are concatenated with the contextual condition and put into the denoiser to recover it from noise. After being denoised for only 50 steps, the tokens are fed into the cVAE decoder to project back to the HOI sequence. The training objective function for the ARDM is
%
\begin{equation}
    \mathcal{L}_{ar} = \frac{1}{K}\sum_{i=1}^{K}||\bm{s_{i}^t} - \bm{\hat{s}}_i(t,c_i)||_2,
\end{equation}
where the $\bm{s}_{i}^t$ is the $i$-th token added with scheduled noise, and $\hat{\bm{s}}_{i}$ is the predicted token at diffusion time step $t$ and contextual condition $c_i$. In the inference phase, the text token encoded by a frozen CLIP text encoder \cite{CLIP} is concatenated with an object point cloud token encoded by the same pre-trained PointNet++ used in cVAE. The Mamba encoder predicts the condition $c_i$ on the $i-$th position, and the MLP denoiser predicts the token with context conditions. In this way, the next token is generated iteratively, until a null token is obtained, which indicates the end of the sequence. We later show in our experiments that our MLP actually outperforms Transformer in accuracy and speed. 

Compared with the previous Text-to-HOI models, our model only uses several MLP layers as denoiser, which can finish denoising in as few as 50 reverse denoising steps using DDIM \cite{ddim}; and the autoregressive network utilizes Mamba instead of a Transformer, which further decreases the inference time.

\noindent \textbf{Classifier-Free Guidance}. To increase the fidelity of the generated model, Classifier-Free Guidance (CFG) \cite{cfg} is applied for training and inference, which has proven to be effective in Text-to-Motion tasks. In the training process, the first text condition $c_{text}$ is sampled as null token with a certain probability. During inference, the denoiser predicts the result with both the text conditions and null conditions, and the final prediction is their combination, 
\begin{equation}
\hat{s}_i^t = \xi \hat{s}_i(z_t, t, c_i) + (1-\xi) \hat{s}_i(z_t, t, c_{i,\emptyset}), 
\end{equation}
where $c_{i,\emptyset}$ is the condition at $i$-th position with empty text token input, and $\xi$ is the CFG factor.

\section{Experiments}

Our experiments are conducted on the OMOMO \cite{omomo}, and BEHAVE \cite{behave} datasets. For the OMOMO dataset, we trim the sequence to a minimum length of 60 and a maximum length of 240 frames. For the BEHAVE dataset, we follow the annotation and sequence splitting by \cite{hoidiff}. The frame rate for both datasets is 30 fps. The canonic motion representation is applied, i.e. the first frame is forced to face the z+ axis. All the human motion samples $\bm{h}_i = \{\bm{t}_i, \bm{r}_i \} \in \mathbb{R}^{T,69}$ are of SMPL representation, composed of root translation and joint rotation in axis angle. The object motions $\bm{o}_i \in \mathbb{R}^{T,6}$ are represented by the combination of object translations and object rotation axis angles in the world system. The two motions are concatenated together as the HOI data representation $\bm{x}_i := \{\bm{h}_i, \bm{o}_i\} \in \mathbb{R}^{T, 75}$ on each frame. For the object point cloud, 256 points are sampled on the objects' surface. 

For evaluation, we follow the common metrics of the Text-to-HOI task \cite{thor}, including Frechet Inception Distance (FID), top-K recall precision (R-precision), Multimodal distance (MMD), and Diversity. For each metric, we repeat the experiments 20 times to avoid coincidence. 

\begin{figure*}
    \centering
    \includegraphics[width=0.9\linewidth]{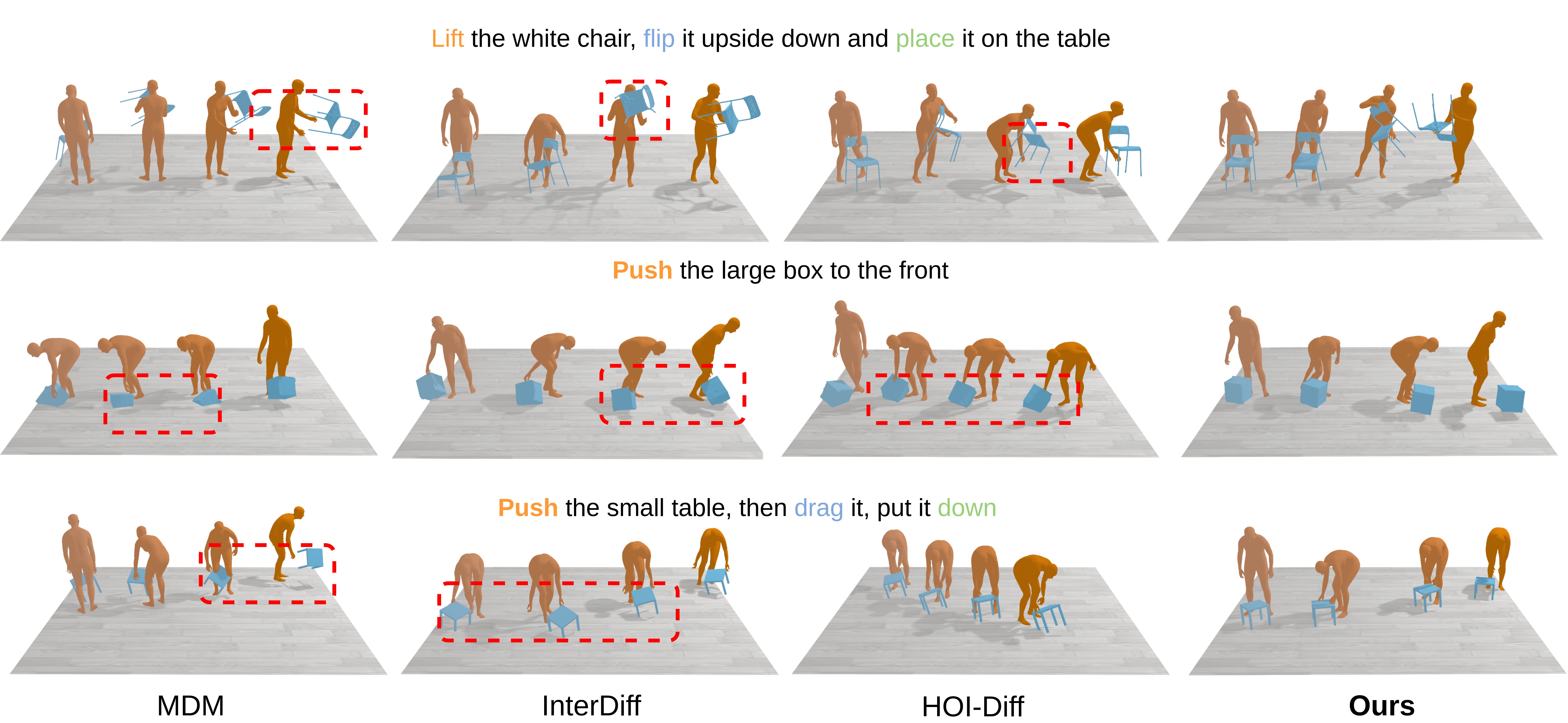}
    \vspace{-2mm}
    \caption{Qualitative comparison with current models. MDM shows strong perturbations. InterDiff and HOI-Diff present a stick-to-hand motion pattern. And their global orientation is not well controlled which leads to penetration. }
    \vspace{-2mm}
    \label{fig:qualitative_cmp}
\end{figure*}


We compare our results with three available architectures MDM \cite{mdm}, InterDiff \cite{hoidiff}, and HOI-Diff \cite{hoidiff}.  MDM and InterDiff are not specifically designed for Text-to-HOI tasks, hence, we modified them for our task (details in Supplementary material). Other works \citet{thor} and \citet{cghoi} have not released their models or code. We contacted them to get their models for comparison, however, we did not receive any response. We tried to implement their methods to reproduce their results, however, due to lack of details, their models could not be trained successfully.


\begin{table}[tp]
\centering
\renewcommand{\arraystretch}{1.02}
\resizebox{\columnwidth}{!}{ 
\begin{tabular}{lccllcc}
\hline
\multicolumn{1}{c}{\multirow{2}{*}{Model}} & \multicolumn{1}{c}{\multirow{2}{*}{FID $\downarrow$}} & \multicolumn{3}{c}{R-precision $\uparrow$}                                              & \multicolumn{1}{c}{\multirow{2}{*}{MMD $\downarrow$}} & \multicolumn{1}{c}{\multirow{2}{*}{ Diversity $\uparrow$}} \\
\multicolumn{1}{c}{}                        & \multicolumn{1}{c}{}                     & top-1                & \multicolumn{1}{c}{top-2} & \multicolumn{1}{c}{top-3} & \multicolumn{1}{c}{}                                     \\
\hline
Ground Truth   &   -  & 0.663  &  0.854   &    0.928       &    5.815   & 6.026 \\ 
Retieval  &  1.216 & 0.647 & 0.858 & 0.927 &  5.809 & 6.008\\
\hline
MDM & 1.392& 0.542 & 0.703 &  0.806 & 5.967 & 5.879\\
InterDiff & 1.268& 0.594 & 0.720 &  0.833 & 5.957 & 5.763\\
HOI-Diff & 1.075& 0.613 & 0.758 &  0.853 & 5.922 & 5.982\\
ARDHOI (ours) & \textbf{0.826}& \textbf{0.628} & \textbf{0.790} &  \textbf{0.884} & \textbf{5.874} & \textbf{6.125}\\
\hline
\end{tabular}
}
\vspace{-2mm}
\caption{Text-to-HOI results on OMOMO dataset. MMD means multimodal distance.}
\label{tab:omomo}
\end{table}

\noindent \textbf{Implementation Details}. In cVAE,  the encoder is a three-block MLP, each consisting of one fully connected layer, a SiLU activation layer, a fully connected layer, and a layer norm. A residual connection is added within the block to avoid model collapse. The decoder has similar settings and is followed by a Mamba layer to learn sequence-wise reconstruction. The channel size of the input is 1024, and the encoded token size is 512. The ARDM consists of 27 Mamba2 layers with a hidden dimension of 512. We use an expansion factor of 2 and the state number is 32. The MLP denoiser has the same setting as the cVAE encoder. Other training settings are discussed in detail in the supplementary material.


\subsection{Quantitive Results for Text-to-HOI} 

Table~\ref{tab:omomo} and \ref{tab:behave} show Text-to-HOI results of our ARDHOI and other available approaches on the OMOMO and BEHAVE datasets. Besides, Table~\ref{tab:speed} suggests a high inference efficiency of our model by comparing the total FLOPs AITS. Our method outperforms the other models on all generative metrics. To illustrate the domain gap between the training and test datasets, we also add a retrieval method, which indices and samples the sequences in the training set with the most similar text descriptions to the samples in the test set.

\begin{table}[tp]
    \centering
    \renewcommand{\arraystretch}{1.02}
    \resizebox{\columnwidth}{!}{ 
    \begin{tabular}{lccllcc}
    \hline
    \multicolumn{1}{c}{\multirow{2}{*}{Model}} & \multicolumn{1}{c}{\multirow{2}{*}{FID $\downarrow$}} & \multicolumn{3}{c}{R-precision $\uparrow$}                                              & \multicolumn{1}{c}{\multirow{2}{*}{MMD $\downarrow$}} & \multicolumn{1}{c}{\multirow{2}{*}{ Diversity $\uparrow$}} \\
    \multicolumn{1}{c}{}                        & \multicolumn{1}{c}{}                     & top-1                & \multicolumn{1}{c}{top-2} & \multicolumn{1}{c}{top-3} & \multicolumn{1}{c}{}                                     \\
    \hline
    Ground Truth   &   -  & 0.715  &  0.847   &    0.900         &    5.515  & 6.417 \\ 
    Retieval  &  1.184 & 0.635 & 0.826 & 0.901 &  5.595 & 6.721\\
    \hline
    MDM & 2.681& 0.310 & 0.448 &  0.581 & 6.014 & 6.382 \\
    
    InterDiff & 2.271& 0.356 &0.511 &  0.608 & 5.905 & 6.471\\
    
    HOI-Diff & 1.998& 0.363 & 0.518 &  0.617 & 5.862 & 6.513\\
    
    ARDHOI (ours)  & \textbf{1.872}& \textbf{0.370} & \textbf{0.525} &  \textbf{0.628} & \textbf{5.835} & \textbf{6.680}\\
    \hline
    \end{tabular}
    }
    \vspace{-2mm}
    \caption{Text-to-HOI results on BEHAVE dataset.}
    \vspace{-4mm}
    \label{tab:behave}
    
\end{table}

\vspace{-2mm}

\begin{table}[tp]
    \centering
    \renewcommand{\arraystretch}{1.0}
    \resizebox{\columnwidth}{!}{ 
    \begin{tabular}{ccccc}
    \hline
    Model  & MDM  & InterDiff & HOI-Diff & ARDM (\textbf{ours}) \\
    \hline
    AITS $\downarrow$   &  29.2 sec &  5.84 sec & 75.3 sec & 1.25 sec \\
    FLOPS $\downarrow$  & 7.36 T  & 1.28 T & 9.57 T & 260.9 B \\
    \hline
    \end{tabular}
    }
    \caption{Inference speed of different architecture, presented by floating operations (FLOPS) and average inference time per sentence in second (AITS).}
    \label{tab:speed}
    
\end{table}
\vspace{-5mm}
\noindent \textbf{Effect of initial state}. In experiments, we found that the quality of the generation result is heavily dependent on the initial state, i.e. the first frames. For the autoregressive model, the first token/frame leads the remained results, and errors get accumulated in the iterative generation. Such a conclusion could be drawn from \cite{chois, interdiff}. Therefore, we add another generation task with both text and the first 16 frames as conditions. The result in Table~\ref{tab:init_omomo} suggests that the autoregressive-based method, i.e. ours and Interdiff have marginally improved performance.
\vspace{-2mm}

\begin{table}[ht]
    \centering
    \renewcommand{\arraystretch}{0.99}
    \resizebox{\columnwidth}{!}{ 
    \begin{tabular}{lccllc}
    \hline
    \multicolumn{1}{c}{\multirow{2}{*}{Model}} & \multicolumn{1}{c}{\multirow{2}{*}{FID $\downarrow$}} & \multicolumn{3}{c}{R-precision $\uparrow$}                                              & \multicolumn{1}{c}{\multirow{2}{*}{MMD $\downarrow$}} \\
    \multicolumn{1}{c}{}                        & \multicolumn{1}{c}{}                     & top-1                & \multicolumn{1}{c}{top-2} & \multicolumn{1}{c}{top-3} & \multicolumn{1}{c}{}                                     \\
    \hline
    Ground Truth   &   -  & 0.663  &  0.854   &    0.928         &    5.815  \\ \hline
    MDM & 1.268& 0.581 & 0.672 &  0.793 & 5.944 \\
    
    InterDiff & 0.801 & 0.656 & 0.802 &  0.905 & 5.858 \\
    
    HOI-Diff & 0.942 & 0.626 & 0.753 &  0.835 & 5.890 \\

    ARDHOI (ours) & \textbf{0.730}& \textbf{0.668} & \textbf{0.827} &  \textbf{0.916} & \textbf{5.807}\\
    \hline
    \end{tabular}
    }
    \caption{Generation results on OMOMO dataset conditioned on text, objects, and initial states.}
    \label{tab:init_omomo}
\end{table}

\vspace{1mm}

\begin{figure}[H]
    \centering
    \includegraphics[width=1.0\linewidth]{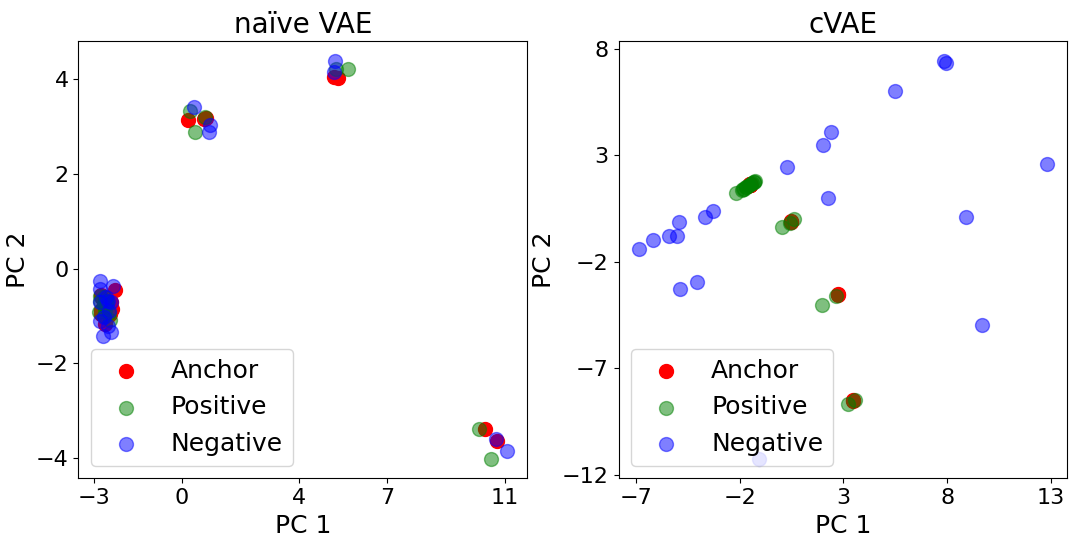}
    \vspace{-3mm}
    \caption{Comparison of PCA plots for HOI positive and negative examples. The plot right is with triplet loss and the plot left is without triplet loss.}
    \label{fig:PCA_Comparison}
    \vspace{-5mm}
\end{figure}

\noindent \textbf{cVAE Space Insight}. To better demonstrate the importance of our cVAE, we present the distance among the anchor samples, positive samples, and negative samples in Fig.~\ref{fig:pipeline} using PCA visualization in Fig.~\ref{fig:PCA_Comparison}. We can see that na\"ive VAE is not able to distinguish physically implausible tokens and the positive and negative tokens both map close to the anchor tokens.  Using the proposed triplet loss, our cVAE pushes the negative tokens away. Our results illustrate the function of our cVAE in blocking implausible tokens. 

\subsection{Qualitative Results}

We present visualizations of model outputs conditioned on textual descriptions and object point clouds from the OMOMO dataset in Fig.~\ref{fig:qualitative_cmp}. Qualitative analysis indicates that our model consistently outperforms the compared methods, generating coherent sequences for both extended and brief interactions. For post-optimized, HOI-Diff and InterDiff, we find that the object and body intended to stick to each other i.e. implausible results. Notably, the animations corresponding to the initial prompt (chairs) in the supplementary materials demonstrate our model's ability to generate Human-Object Interaction (HOI) sequences, even in scenarios involving rapid rotational movements.

\subsection{Ablation Study}

\noindent \textbf{Token frames.} The 
number of frames/tokens plays an important role in the generation, so we experimented with different token sizes. Table~\ref{tab:framesize} and qualitative results attached in the supplementary indicate that our method achieves optimal performance when the token size is 16 frames i.e. 0.5 seconds sequence at 30fps. 

\begin{table}[htp]
    \renewcommand{\arraystretch}{1.0}
    \centering
    \resizebox{\columnwidth}{!}{ 
    \begin{tabular}{cccllc}
    \hline
    \multicolumn{1}{c}{ARDHOI} & \multicolumn{1}{c}{\multirow{2}{*}{FID $\downarrow$ }} & \multicolumn{3}{c}{R-precision $\uparrow$} & \multicolumn{1}{c}{\multirow{2}{*}{MMD $\downarrow$}} \\
    \multicolumn{1}{c}{Token size}                        & \multicolumn{1}{c}{}                     & top-1                & \multicolumn{1}{c}{top-2} & \multicolumn{1}{c}{top-3} & \multicolumn{1}{c}{}                                     \\
    \hline
    Ground Truth &   -  & 0.663  &  0.854   &    0.928         &    5.815   \\ \hline
    1  & 1.289 & 0.593 & 0.738 &  0.843 & 5.921\\
    2  & 1.267 & 0.608 & 0.744 &  0.851 & {5.955}\\
    4  & 1.306 & 0.614 & 0.759 &  0.862 & {5.932}\\
    8  & 1.023 & 0.622 & 0.775 &  0.871 & {5.880}\\
    10  & 0.929 & 0.621 & 0.780 &  0.870 & {5.883}\\
    12  & 0.886 & 0.616 & \textbf{0.794} &  {0.874} & {5.879}\\
    16  & \textbf{0.826} & \textbf{0.628} & {0.790} &  \textbf{0.884} & \textbf{5.874}\\
    20  & 0.977 & 0.619 & 0.779 &  {0.870} & {5.877}\\
    24  & 1.084 & 0.608 & 0.757 &  0.864 & {5.901}\\
    
    \hline
    \end{tabular}
    }
    \caption{Ablation study for Text-to-HOI on the OMOMO dataset with token sizes from 1 to 24.}
    \label{tab:framesize}
\end{table}

\noindent \textbf{Contrastive Loss}. To demonstrate the importance of our cVAE, we conducted an ablation study by removing the contrastive loss component. The quantitative results in Table~\ref{tab:ablation} and Table~\ref{fig:PCA_Comparison}, reveal a marked decline in performance, highlighting the critical role of contrastive loss in our model.

\noindent \textbf{Mamba Context Encoder} not only achieves fast inference speed but also achieves better generative performance as indicated in Table \ref{tab:ablation}. Additionally, visualization results show that Mamba has a natural 
smoothing effect. 

\noindent \textbf{MLP Denoiser}. We examine the function and efficacy of our MLP denoiser by replacing it with a Transformer Encoder. Table~\ref{tab:ablation} suggests that compared to a Transformer-based Denoiser, our MLP denoiser is more capable of denoising in our autoregressive paradigm. 

\noindent \textbf{Diffusion Loss}. The diffusion process serves an important role in token prediction. To verify the efficacy of the inverse diffusion process, we skip the iteration of inverse diffusion and optimize the MSE loss of predicted tokens and the ground truth tokens. Table~\ref{tab:ablation} shows that this significantly degrades the results.

\begin{table}[htp]
\centering
\renewcommand{\arraystretch}{1.}
\resizebox{\columnwidth}{!}{ 
\begin{tabular}{llcllcc}
\hline
    \multicolumn{1}{c}{\multirow{2}{*}{Model}} & \multicolumn{1}{c}{\multirow{2}{*}{FID $\downarrow$}} & \multicolumn{3}{c}{R-precision $\uparrow$}                                              & \multicolumn{1}{c}{\multirow{2}{*}{MMD $\downarrow$}} & \multicolumn{1}{c}{\multirow{2}{*}{AITS $\downarrow$}}\\
    \multicolumn{1}{c}{}                        & \multicolumn{1}{c}{}                     & top-1                & \multicolumn{1}{c}{top-2} & \multicolumn{1}{c}{top-3} & \multicolumn{1}{c}{}    & \multicolumn{1}{c}{}                                 \\
\hline
Ground Truth  & - & 0.663  &  0.854   &    0.928       &    5.815  & - \\ \hline
w.o. Triplet Loss & 0.948 & 0.618 & 0.763 &  0.860 & 5.892 & 1.252 \\
TRM Context Encoder & 0.979 & 0.616 & 0.768 &  0.850 & 5.886 & 1.790 \\
TRM Denoiser & 0.902 & 0.620 & 0.787 &  0.875 & 5.875 & 3.571\\
MSE Loss & {1.288}& {0.535} & {0.704} &  \textbf{0.810} & {5.967} & 0.625\\
Full & \textbf{0.826}& \textbf{0.628} & \textbf{0.790} &  \textbf{0.884} & \textbf{5.874} & 1.252 \\
\hline
\end{tabular}
}
\caption{Ablation study. The first row w.o. triplet loss removes the triplet loss in the cVAE. The TRM Context Encoder replaces the Mamba blocks with Transformer Encoder Blocks. The TRM Denoiser replaces the MLP denoiser with a Transformer Decoder (cross-attention). The L2 Loss replaces the denoise process and explicitly predicts the next token with MSE loss in one step.}
\label{tab:ablation}
\end{table}

\section{Conclusion}

We proposed ARDHOI, a pipeline for generating 3D Human-Object Interaction sequences driven by text. Our method first tokenizes the small HOI intervals using a Contrastive Variational Autoencoder to learn a physically plausible continuous latent space. Using contrastive learning, we increase the margin between plausible and implausible HOIs. To operate autoregressive generation on such a continuous token space, we proposed an Autoregressive Diffusion Model based on the Mamba context encoder and MLP denoiser. The Mamba encoder encodes the contextual information and passes the contextual condition to guide the MLP denoiser. The HOI token is then fed to the VAE decoder to project back to the HOI sequence. Extensive experiments on two benchmark datasets and ablation studies show the efficacy of our proposed method.


\section{Acknowledgments}
This work was supported by the Australian Research Council Industrial Transformation Research Hub IH180100002. Professor Ajmal Mian is the recipient of an Australian Research Council Future Fellowship Award (project number FT210100268) funded by the Australian
Government.

\bibliography{aaai25}

\end{document}